# An Audit Framework for Technical Assessment of Binary Classifiers

Debarati Bhaumik[1][a] and Diptish Dey[1][b]
[1]*Amsterdam University of Applied Sciences, the Netherlands*
{d.bhaumik, d.dey2}@hva.nl

Keywords: Auditable AI, Multilevel Logistic Regression, Random Forest, Explainability, Discrimination, Ethics

Abstract: Multilevel models using logistic regression (MLogRM) and random forest models (RFM) are increasingly deployed in industry for the purpose of binary classification. The European Commission's proposed Artificial Intelligence Act (AIA) necessitates, under certain conditions, that application of such models is fair, transparent, and ethical, which consequently implies technical assessment of these models. This paper proposes and demonstrates an audit framework for technical assessment of RFMs and MLogRMs by focussing on model-, discrimination-, and transparency & explainability-related aspects. To measure these aspects 20 KPIs are proposed, which are paired to a traffic light risk assessment method. An open-source dataset is used to train a RFM and a MLogRM model and these KPIs are computed and compared with the traffic lights. The performance of popular explainability methods such as kernel- and tree-SHAP are assessed. The framework is expected to assist regulatory bodies in performing conformity assessments of binary classifiers and also benefits providers and users deploying such AI-systems to comply with the AIA.

## 1 INTRODUCTION

The large-scale proliferation of AI/ML systems within a relatively short span of time has been accompanied by their undesirable impact on society (Angwin, Larson, Mattu, & Kirchner, 2016). For example, it takes the forms of price discrimination in online retail based upon geographic location (Mikians, Gyarmati, Erramilli, & Laoutaris, 2012), and in unsought societal impact of algorithmic pricing in tourism and hospitality (van der Rest, Sears, Kuokkanen, & Heidary, 2022). Often deploying classification models these systems, due to their increasing complexity, are often uninterpretable by humans. The mounting struggle to fully comprehend the rationale behind decisions made by these systems (Biran & Cotton, 2017), also fuels the need to examine and scientifically explain such rationale (explainability) (Miller, 2019) and approach explainability in a more structured and actionable manner (Raji & Buolamwini, 2019; Kazim, Denny, & Koshiyama, 2021). The need for explainability is driven among others, by discrimination/biasness concerns of individuals, ethical considerations of psychologists and social activists, corporate social responsibility objectives of corporations and the responsibilities of legislators to protect fundamental rights of citizens (Schroepfer, 2021; Kordzadeh & Ghasemaghaei, 2022; Rai, 2022).

Technically assessing AI systems and making them explainable includes activities such as validating model assumptions, or generating model explanations (Miller, 2019). Such activities are very specific to the choice of algorithms and underlying model assumptions. Recognizing the importance of technology-specific approach, our proposed framework focusses on binary classification problems. Commonly applied binary classification models (algorithms) such as logistic regression (LR) and random forest (RF) are selected. Owing to their simplicity and their complementary scopes of application, these algorithms are heavily deployed (Beğenilmiş & Uskudarli, 2018). LR, when combined with multilevel models (Gelman & Hill, 2006), extends the performance of the former considerably. These MLogRMs are beneficial in circumstances, where a combination of local behaviour and global context needs to be accounted for, such as applications within groups or hierarchies. Our proposed audit framework is developed for application within binary classification problems.

The paper is organized as follows. Section 2 presents a short summary of the AIA. Section 3

---

[a]  https://orcid.org/0000-0002-5457-6481
[b]  https://orcid.org/0000-0003-3913-2185

presents an audit framework applicable to two classification models, MLogRM and RF, from a theoretical perspective highlighting characteristics that potentially play a role in their audit. Section 5 demonstrates the framework using an online open dataset. Finally, section 6 presents conclusions with a view to the future.

## 2 THE PROPOSED EU AI ACT

Recognizing the roles and concerns of a wide range of stakeholders, the European Parliament proposed the AIA (Commissie, 2021), which presents a conformity regime. Using a tiered risk-based approach, the AIA defines three tiers, "low- or no-risk", "high-risk" and "unacceptable risk"; page 12 of the Explanatory Memorandum of the AIA defines these tiers. High-risk systems are subject to ex-ante conformity assessments and post-market monitoring systems driven by competence in specific AI technologies. Although the AIA recognises the importance of competence, it does not elaborate on the role competence plays in specific AI technologies; albeit, "expertise" constitutes one of the 5 key criteria of a good regulation (Baldwin, Cave, & Lodge, 1999). Furthermore, distinction is made between two types of high-risk AI systems: stand-alone high-risk AI systems and those that function as components in consumer products. The latter are subject to legislation specific to their inherent sector. On the former, the AIA stipulates that a provider of a stand-alone high-risk AI system can choose to conduct ex-ante conformity assessment internally if its AI system is fully compliant or use an external auditor if its AI system is either partially compliant or harmonized standards governing that AI system is non-existent.

The AIA addresses accountability in the supply chain and focusses also on AI/ML models that process data. Aiming to tackle the issues of "remote biometric identification" and "biometric categorisation", the AIA encompasses various standpoints, among others, prohibited practices, transparency obligations, governance, and compliance procedures across the supply chain of these systems (Bhaumik, Dey, & Kayal, 2022). It "*sets out the legal requirements for high-risk AI systems in relation to data and data governance, documentation and recording keeping, transparency and provision of information to users, human oversight, robustness, accuracy and security*" (Commissie, 2021, p. 13). Articles 10 to 15 of Chapter 2 Title III of the AIA elaborates on these legal requirements.

As legislations evolve towards regulatory frameworks and eventually result in enforcement strategies such as in the AIA, the mechanisms deployed in enforcement play a critical role in the eventual success of the legislation (Baldwin et al, 1999). In the AIA regime, enforcement activities would include maintaining registers of high-risk AI systems and technical audit of such systems.

## 3 AUDIT FRAMEWORK FOR TECHNICAL COMPLIANCE

Binary classification is one of the most ubiquitous tasks required to solve problems in almost every industry. From default prediction in credit risk to diagnosing benign or malignant cancer in healthcare to churn prediction in telecom, binary classifiers are heavily used for decision making (Kim, Cho, & Ryu, 2020; Esteva, Kupre, Novoa, Ko, Swetter, Blau, & Thrun, 2017). The task of a binary classifier is to predict the class (1/0 or yes/no) to which a particular instance belongs, given some input features, e.g., to predict if an obligor will default on their loan given her credit rating, marital status, income, and gender. With the advent of higher computing power, AI/ML systems are increasingly deployed for making binary classification decisions in different domains (Xu, Liu, Cao, Huang, Liu, Qian, Liu, Wu, Dong, Qiu, & Qiu, 2021; Sarker, 2021). There exists many AI/ML algorithms of varying complexity to perform binary classification tasks. Amongst the more commonly used AI/ML classifiers, LRs, decision trees, K-nearest neighbors, and MLogRMs are less complex to construct, whereas RFs, support vector machines, XGBoost, artificial neural networks, and deep neural networks have more complex architectures. The higher the model complexity, the higher its prediction accuracy. However, this comes at the cost of explainability of model predictions (Gunning, Stefik, Choi, Miller, Stumpf, & Yang, 2019).

### 3.1 Common Binary Classifiers

LR and RF are two commonly used classifiers for both binary and muti-class classification tasks in different sectors such as finance, healthcare, agriculture, IT and many more. Applications range from credit scoring in finance to diabetes detection in healthcare, forest fire and soil erosion prediction in agriculture, cyber security intrusion detection in IT (Dumitrescu, Hué, Hurlin, & Tokpavi, 2022; Daghistani & Alshammari, 2020; Milanović, Marković, Pamučar, Gigović, Kostić, & Milanović, 2020; Ghosh & Maiti, 2021; Gupta & Kulariya, 2016). For the majority of such applications, it is observed that RF classifiers tend to achieve higher accuracy compared to LR classifiers. However, this high accuracy comes at the cost of explainability of the classifier's predictions. Therefore, in highly regulated sectors, such as banking, LR classifiers are still preferred over the more accurate but more 'opaque' RF classifiers (EBA, 2020).

When a dataset has inherent group or hierarchical structure, the accuracy of LR classifiers can be increased without losing their explainability power, by coupling them with multilevel models (MLMs), resulting in MLogRMs (Gelman & Hill, 2006). MLogRMs are capable of capturing the individual level (local) effects while preserving the overarching group level (global) effects. MLogRMs are used increasingly in industry across different sectors such as insurance (Ma, Baker. & Smith, 2021), healthcare (Adedokun, Uthman, Adekanmbi, & Wiysonge, 2017), agriculture (Giannakis & Bruggeman 2018, Nawrotzki & Bakhtsiyarava, 2017), and urban planning (Wang, Abdel-Aty, Shi, & Park, 2015).

Given the high penetration of binary classifiers in real-world industrial applications, it is important that these classifiers are implemented in a trustworthy manner. Fairness and transparency are two commonly used attributes of a trustable system/model. Fairness can be achieved through detection and avoidance of biases in these systems, which can otherwise, potentially lead to discrimination of minorities and unprivileged groups (Mehrabi, Morstatter, Saxena, Lerman, & Galstyan, 2021). Bias could creep into the classifiers in three phases of their model lifecycle: (i) pre-processing phase: when model training data is collected, (ii) in-processing phase: when the algorithm of the classifier is being trained on the collected data, and (iii) post-processing phase: when the model is in production and is not monitored for model re-training. Therefore, it is not only important to have technical expertise for bias detection, but also process expertise, to ensure procedures are in place for bias mitigation. Transparency is the other attribute of a trustable system. In case of binary classifiers this implies that the predictions made by the classifier can be explained to various stakeholders, such as, model developers, model users, and end users.

This paper is incremental and builds upon the audit framework applied to MLogRM by Bhaumik et al (2022). For the sake of completeness of the framework, this paper repeats some of their conclusions. This paper is innovative in two ways: firstly, it extends their audit framework and introduces additional sub-aspects such as *robustness* and *accuracy of counterfactual explanations* and secondly, it applies the audit framework to go beyond MLogRMs and include RFMs.

### 3.2 Multilevel Logistic Regression

As stated earlier, MLogRMs are used in binary classification problems where data is structured in groups or is characterized by inherent hierarchies. MLogRMs can be perceived as a collection of multiple LR models that vary per MLM group. These multiple LR models are modeled using random variables with a common distribution. Therefore, the model parameters (coefficients) of MLogRMs vary per MLM group but they come from a common distribution. The common distribution of the model parameters leads to higher predictive power compared to simple LR models. MLogRMs come in three flavors: (i) random intercept model, where only the intercepts vary per MLM-group, (ii) random slope model, where only the slopes vary per MLM-group, and (iii) random intercept and slope model, where both the intercepts and the slopes vary per MLM-group (Gelman and Hill, 2006). Equation 1 presents of a simple random intercept and slope MLogRM model with three independent variables $x^1, x^2, x^3$ and a dependent variable $y$:

$$\log\left(\frac{\mathbb{P}(y_i = 1)}{\mathbb{P}(y_i = 0)}\right) = \alpha_{[j]i} + \beta^1_{[j]i} x^1_i + \beta^2_{[j]i} x^2_i + \beta^3_i x^3_i + \epsilon_i. \quad (1)$$

In equation (1), $i = \{1, ..., N\}$, $N$ being the total number of data points, $j = \{1, ..., J\}$, $J$ being the number of MLM-groups. Model assumptions represented in equation (1) include:

- $\log\left(\frac{\mathbb{P}(y_i = 1)}{\mathbb{P}(y_i = 0)}\right)$, the log-odds term, has a linear relationship with the independent variables $x_i$'s.
- $x's$, the independent variables, are mutually uncorrelated.
- $\epsilon'_i s$, the error terms, are normally distributed with mean zero.
- The varying intercept and slope terms follow normal distributions.
- The training and testing datasets follow the same distribution.

These assumptions lead to the following *statistical properties* for the model in equation (1):

- $\epsilon_i \sim \mathcal{N}(0, \sigma^2_y)$,
- $\alpha_{[j]} \sim \mathcal{N}(\mu_\alpha, \sigma^2_\alpha)$, and $\beta_{[j]} \sim \mathcal{N}(\mu_\beta, \sigma^2_\beta)$, for $j = 1, ..., J$.
- $D_{training} \triangleq D_{test}$.

MLogRMs can only capture linear relationship between independent and dependent variables. Therefore, predictive power of these models decreases when there exists a non-linear relationship between independent and dependent variables.

Table 1: Audit framework for technical assessment.

| Aspects | Sub-aspects | Contributors | KPIs |
|---|---|---|---|
| 1. Model<br>*Assess that a model fitted to data is technically stable and valid.* | 1.1 Formulation of relevant assumptions<br>*Tests are performed to ensure validity of model assumptions and statistical properties.* | Model assumptions | 1.1.1 Presence in technical documentation |
| | | Statistical properties | 1.1.2a VIF |
| | | | 1.1.2b SWT |
| | | | 1.1.2c BPT |
| | 1.2 Accuracy of predictions<br>*Checks are undertaken to monitor accuracy of model predictions.* | Discriminatory power | 1.2.1 AUC-ROC |
| | | Predictive power | 1.2.2 *F1*-score |
| | 1.3 Robustness<br>*Tests are conducted to assess sensitivity of model output.* | Sensitivity of model output to change in model parameter | 1.3.1 TSVR |
| | | Sensitivity of input around inflection points | 1.3.2 CSVP |
| 2. Discrimination<br>*Assess that a model is fair at individual and group levels.* | 2.1 Group fairness<br>*Tests are performed to check if group fairness for equal group treatment is within an acceptable range.* | Predicted versus actual outcome | 2.1.1 EqualOdds |
| | | Predicted equality | 2.1.2a DI |
| | | | 2.1.2b SP |
| | 2.2 Individual fairness<br>*Tests are performed to check if individual fairness for equal treatment of individuals is within an acceptable range.* | Intra-group fairness | 2.2.1 $\text{Diff}_{Ind}$ |
| | | Inter-group fairness | 2.2.2 $\text{Diff}_{Ind\_GRP}$ |
| 3. Transparency & explainability<br>*Assess the extent to which model predictions are explainable to humans and suggest actions that would facilitate a (alternative) desirable prediction.* | 3.1 Accuracy of feature attribution explainability methods<br>*Tests are performed to assess quality of feature attribution explainability methods used to explain model predictions to humans.* | Feature contribution order | 3.1.1 $\rho_{order}$ |
| | | Aggregated feature contribution<br>*(because individuals cannot be compared)* | 3.1.2 PUX |
| | | Feature contribution sign | 3.1.3 POIFS |
| | 3.2 Accuracy of counterfactual explanations<br>*Tests are performed to assess quality of counterfactuals generated to substantiate a model output.* | Percentage of valid counterfactuals | 3.2.1 PVCF |
| | | Proximity | 3.2.2 PCF |
| | | Sparsity | 3.2.3 SCF |
| | | Diversity | 3.2.4 DCF |

## 3.3 Random Forests

RFMs are known to have higher accuracy and predictive power when compared to LR models. This is because RFMs can learn non-linear relationships between the independent and the dependent variables (Breiman, 2001). A RFM is an ensemble of decision trees build from different bootstrapped samples of the training dataset, along with taking a random subset of the independent variables for each decision tree. A decision tree is a supervised learning method where the feature space is partitioned recursively into smaller regions. The recursive partition of the feature space is done through a set of decision rules which typically results in 'yes/no' conditions. These cascading 'yes/no' decision rules can be seen as a tree, hence the name *decision tree*. The class which gets the most votes from the ensemble of decision trees is the final predicted class by the RFM.

The ensembles of decision trees help in reducing the variance of the classifier, resulting in higher predictive power. The widespread use of RFMs in industrial applications is not only due to their high predictive power but also due to their inbuilt global *variable importance measure* (VIM) for the independent/feature variable. This makes RFMs more explainable at the global level unlike other 'black-box' models. However, the global VIM fails to explain local or instance-based predictions. A new methodology, dimensional reduction forests, formulated by Loyal, Zhu, Cui, & Zhang (2022) can provide explanations for local predictions. However, the method is not performance tested against the commonly used feature attribution explanation methods SHAP and LIME (Molnar, 2020).

Being non-parametric in nature, RFMs do not have any distributional assumptions. However, for unbiased model performance metrics calculations, it is important that both the training and test datasets follow the same distribution i.e., $D_{training} \triangleq D_{test}$.

## 3.4 Framework for Technical Audit of Classification Models

As discussed in the sections 3.2 and 3.3, technical assessment of binary classifiers would require assessing various algorithmic aspects. These include: (i) *model aspects*, (ii) *discrimination aspects*, and (iii) *transparency & explainability aspects*, which are presented in our proposed audit framework (Table 1). The following sections of this paper present various sub-aspects, contributors and their corresponding KPIs for assessing these algorithmic aspects.

### 3.4.1 KPIs for Model Assumptions and Statistical Properties

The KPI associated with assessing *model assumption*s is qualitative as it is assessed by the extent of their presence in technical documentation. The KPIs for assessing *statistical properties* presented in Table 1 are variance inflation factor (*VIF*), Shapiro-Wilk test (*SWT*) and Breusch-Pagan test (*BPT*). VIF measures the presence of multi-collinearity among independent variables and a value of VIF > 5.0 should be mitigated. SWT is a statistical test which is used to check normality in data. For MLogRM this test is applied to check if model residuals are normally distributed. Results of MLogRM can be trusted only if residuals of a fitted model have constant variances. This assumption is checked using BPT that checks for heteroscedasticity in data samples. For both SWT and BPT if *p-values* are close to zero, then the null hypothesis of normality and constant variance in sample data is rejected in favour of the alternative hypothesis.

### 3.4.2 KPIs for Accuracy of Predictions

The KPIs associated with measuring the sub-aspect *accuracy of predictions* are Area under the ROC curve (*AUC-ROC*) and *F1*-score. *F1*-score measures the predictive power and is the geometric mean of precision and recall of a binary classifier. Whereas, *AUC-ROC,* measures how good a binary classifier is in separating the two classes (i.e., the discriminatory power). Both of these KPIs take values between [0,1] and the closer their values are to 1, the more accurate the binary classifier is.

### 3.4.3. KPIs for Robustness

To assess the *robustness* of a model, two contributors have been proposed, *sensitivity of the model output to change in model parameter* and *sensitivity of input around the inflection points*. The KPIs proposed to measure these contributors are total Sobol's variance ratio (*TSVR*) and cosine similarity vector pairs (*CSVP*) respectively.

*TSVR* is computed by taking the sum of the first-order sensitivity indices over all the estimated model parameters. The first-order sensitivity index, also known as the Sobol's variance ratio for the $i^{th}$ parameter, is a ratio of the variance of a model output under the variation of a single model parameter (for example $\beta_i$ in equation (1)) to the variance of the model output (Tosin, Côrtes, & Cunha, 2020). Mathematically,

$$S_i = \frac{Var_{\beta'_j}(Y|\boldsymbol{\beta}_{i \neq j}, \overline{\boldsymbol{X}}_{test})}{Var(Y|\boldsymbol{\beta}, \boldsymbol{X}_{test})}, \quad (2)$$

where, $Var_{\beta'_j}(Y|\boldsymbol{\beta}_{i \neq j}, \overline{\boldsymbol{X}}_{test})$ is the variance of the model output in which the $j^{th}$ parameter is varied while keeping the other model parameters $\boldsymbol{\beta}_{i \neq j}$ at constant and the input independent variables/ feature values of the test dataset ($\boldsymbol{X}_{test}$) at their mean values, i.e., $\overline{\boldsymbol{X}}_{test}$. $Var(Y|\boldsymbol{\beta}, \boldsymbol{X}_{test})$ is the variance of the model output on the test dataset $\boldsymbol{X}_{test}$ and $\boldsymbol{\beta}$ are the estimated model parameters.
Finally,

$$TSVR = \sum_{i=1}^{P} S_i, \quad (3)$$

where, *P* is the total number of estimated model parameters. For the MLogRM in the case-study there are 4 model parameters, one corresponding to the intercept term $\alpha$ in equation (1) and the other three corresponding to the slope terms from age, BMI and number of children, i.e., $\beta^1, \beta^2$ and $\beta^3$, respectively. We vary the $j^{th}$ model parameter $\beta'_j$ between the interval $[\beta_j - SE_{\beta_j}, \beta_j + SE_{\beta_j}]$, where $SE_{\beta_j}$ is the standard error of the estimated model parameter $\beta_j$.

A smaller value of *TSVR* corresponds to a better model, because the model then is less sensitive under the variation of parameters. The value of *TSVR* has an upper bound of 1 – high order interaction terms. The higher order interaction terms are non-trivial, therefore knowing the exact upper bound of *TSVR* is difficult, however it cannot exceed the value of 1.

*CSVP* is the number of input vector pairs predicted to be in different classes around the inflection point ($p = 0.5$) with cosine similarity greater that $1 - \delta$. The inflection point can be seen as the probability threshold point to classify an input data point if it belongs to class 0 (for $p < 0.5$) or class 1 (for $p \geq 0.5$). In simpler terms, *CSVP* finds the number of vector pairs that are very similar in values but have been predicted to be in different classes. This metric helps in assessing how sensitive are the model predictions for very similar datapoints around the point of inflection. For computing *CSVP*, a neighborhood of $[p - 0.01, p + 0.01]$ was considered around the inflection point and $\delta = 0.1$ was taken for cosine similarity. Therefore, only those vector pairs were chosen around the inflection point, which had cosine similarity greater than 0.9 and were predicted to be in different classes. A lower value of *CSVP* results in a more robust classifier.

### 3.4.4 KPIs for Discrimination

As proposed by Pessach & Schmueli (2022) and Hardt, Price, & Srebro (2016), discrimination can be assessed through two sub-aspects, *group fairness* and *individual fairness*. The KPIs proposed to assess *group fairness* are *Statistical Parity (SP)*, *Disparate Impact (DI)* and *EqualizedOdds (EqualOdds)*. *SP* and *DI* measure the difference between the positive predictions across the sensitive groups $S$, such as gender or ethnicity. These KPIs can be calculated as:

$$SP = \left| \mathbb{P}(\hat{Y} = 1 | S = 1) - \mathbb{P}(\hat{Y} = 1 | S \neq 1) \right| \quad (4)$$

and,

$$DI = \frac{\mathbb{P}[\hat{Y}=1|S=1]}{\mathbb{P}[\hat{Y}=1|S\neq 1]} \quad (5)$$

In the equations above, $\hat{Y} = 1$ represents the favourable class, $S$ represents the feature or attribute that is possibly discriminatory, and $S \neq 1$ represents the under-privileged group. A low value of *SP* in equation 4 and a high value of *DI* in equation 5 imply that a favoured classification is similar across different groups. One of the few available benchmarks for these KPIs is that of an acceptable value of *DI* of 0.8 or higher (Stephanopoulos, 2018).

Equalized odds, the mean of differences between false-positive rates and true-positive rates, is given by

$$EqualOdds = \frac{1}{2}(Diff_{FPR} + Diff_{TPR}), \quad (6)$$

where, $Diff_{FPR}$ is calculated as the absolute value of the difference between $\mathbb{P}[\hat{Y} = 1 | S = 1, Y = 0]$ and $\mathbb{P}[\hat{Y} = 1 | S \neq 1, Y = 0]$. $Diff_{TPR}$ is calculated as the absolute value of the difference between $\mathbb{P}[\hat{Y} = 1 | S = 1, Y = 1]$ and $\mathbb{P}[\hat{Y} = 1 | S \neq 1, Y = 1]$. Ideally *EqualOdds* should be zero.

The KPIs proposed to assess *individual fairness* are $Diff_{Ind}$ and $Diff_{Ind\_GRP}$ which checks if similar individuals are treated equally within a MLM-group and across MLM-groups respectively. $Diff_{Ind}$ is calculated using the formula

$$\begin{aligned} Diff_{Ind} &= \left| \mathbb{P}(\hat{Y}^{(i)} = y | X^{(i)}, S^{(i)}) - \mathbb{P}(\hat{Y}^{(j)} = y | X^{(j)}, S^{(j)}) \right|, \\ & \text{if } d(i,j) \approx 0 \end{aligned} \quad (7)$$

where, $i$ and $j$ denote two individuals with $S$ and $X$ representing sensitive attributes and associated features of the individuals respectively, $d(i,j)$ is the distance metrics and its value close of zero ensures that the two individuals compared are in some respect similar. $Diff_{Ind\_GRP}$ is calculated by

$$\begin{aligned} Diff_{Ind\_GRP} &= \left| \mathbb{P}(\hat{Y}^{(i,a)} = y | X^{(i,a)}, S^{(i,a)}) \right. \\ & \quad - \mathbb{P}(\hat{Y}^{(j,b)} \\ & \quad \left. = y | X^{(j,b)}, S^{(j,b)}) \right| \end{aligned} \quad (8)$$

where $a$ and $b$ refer to two different MLM-groups. The smaller the values of these KPIs, the fairer similar individuals are treated by the classifier.

### 3.4.5 KPIs for Transparency and Explainability

Predictive power and accuracy of classification models are inversely proportional to their complexity. The more complex a model gets the more difficult it becomes to explain and understand outputs. Therefore, for trustable use of AI classifiers at scale, it is important that these classifiers are algorithmically transparent, and that their predictions are explainable (Chen, Li, Kim, Plumb, & Talwalkar 2022; Dwivedi, Dave, Naik, Singhal, Rana, Patel, Qian, Wen, Shah, Morgan, & Ranjan, 2022). *Transparency and explainability* of AI classifiers can be achieved through comprehensibility of the underlying model decision making process and providing human-understandable explanations for model predictions (Mohseni, Zarei, & Ragan, 2021).

Two commonly used post-hoc explanation methods are *feature attribution explanations* and *counterfactual explanations*. Feature attribution explanations explicates how much each feature of an input datapoint contributed to the model prediction, whereas *counterfactual explanations* provide actionable alternatives that will lead to desired predicted outcomes (Molnar, 2020; Mothilal, Sharma, & Tan, 2020).

The proposed sub-aspects to assess the transparency and explainability are: (i) *accuracy of feature attribution explainability methods* and (ii) *accuracy of counterfactual explanations.*

**KPIs for Accuracy of Feature Attribution Explanations**

Feature attribution explanations elucidate the importance of each feature by calculating feature contribution to model prediction. Two commonly used methods in industry are SHAP (SHapley Additive exPlanations) and LIME (local linear regression model) (Loh, Ooi, Seoni, Barua, Molinari, & Acharya, 2022).

The proposed contributors to assess the accuracy of feature attribution explainability methods include *feature contribution order*, *aggregated feature contribution*, and *feature contribution sign*. These contributors are assessed through comparing SHAP explanations with model intrinsic explanation methods. KPIs proposed to assess the above-mentioned contributors are *Spearman's rank correlation coefficient* ($\rho_{order}$), *probability unexplained* ($PUX$) and *percentage of incorrect feature signs* ($POIFS$).

For MLogRMs, the model intrinsic feature attribution explanations can be generated using the estimated model parameters in equation (1). The log-odds term has a linear relationship with the independent (feature) variables (the $x's$); the magnitude and sign of the $\beta's$ represent the change of log-odds of an individual being predicted for class 1 when the input feature values (the $x's$) are increased by one unit. Similarly, the $\alpha's$ are the base (reference) value contribution to the log-odds term when all the feature variables are set to zero. Therefore, for any data instance, the contribution of each feature to the log-odds of an individual prediction is easily calculated.

For RFM classifiers, the most commonly used model intrinsic feature attribution explanations are generated by calculating *Gini impurity/entropy* from their structure, also known as variable importance measure (VIM) (Strobl, Boulesteix, Kneib, Augustin, & Zeileis, 2008). The decision nodes of RFMs are partitions in the feature space which are generated by calculating if the decision of partitioning a feature has reduced the Gini impurity or increased the entropy at the node. Feature importance per decision tree is measured by finding how much each feature contributed to reducing the Gini impurity. VIM for an RFM is computed by taking the average of the feature importance over the set of the decision trees it is made of. It is important to note that VIM is a global explanation method which does not provide feature attribution explanations for individual data instances.

The SHAP method, based on Shapely values from cooperative game theory, can produce both local (instance based) and global explanations based on feature attribution magnitude. Lundberg & Lee (2017) have developed model agnostic and model specific SHAP methods beside creating Python libraries for computing SHAP values. For MLogRMs model agnostic *kernel*-SHAP is used and for RFMs model specific *tree*-SHAP is used in this paper.

The proposed KPIs are defined as:

- $\rho_{order}$ is computed by comparing the ranks of feature contribution magnitude of SHAP with model intrinsic method. The values of $\rho_{order}$ range between [-1, 1], with $-1$ being the worst-case negative association and $+1$ being the best-case positive association.
- $PUX$ is the absolute value of the difference between the probability estimates obtained from the model intrinsic method, $\mathbb{P}(y_i = 1)_{MI}$ and those obtained from the SHAP, $\mathbb{P}(y_i = 1)_{SHAP}$, i.e., $|\mathbb{P}(y_i = 1)_{MI} - \mathbb{P}(y_i = 1)_{SHAP}|$ . The ideal value for $PUX$ is zero.
- $POIFS$ is the ratio of the number of times SHAP estimate the feature contribution sign incorrectly as compared to the model intrinsic method with the total number of features, i.e., $\frac{\#incorrect\ feature\ signs}{\#total\ features} \times 100$. The ideal value for $POIFS$ is zero.

**KPIs for Accuracy of Counterfactual Explanations**

The proposed contributors for assessing the accuracy of counterfactual explanations are *validity, proximity, sparsity* and *diversity* of counterfactual explanations generated (Mothilal et al, 2020). The KPIs associated with these contributors are *percentage of valid counterfactuals (PVCF), proximity of counterfactuals (PCF), sparsity of counterfactuals (SCF),* and *diversity of counterfactuals (DCF).*

Given a set $C = \{c_1, c_2, \ldots, c_k\}$ of $k$ counterfactual examples generated for an original input $x$, following are the KPI definitions:

- $PVCF$ computes the % of times generated counterfactuals are actually counterfactuals, i.e.,

$$PVCF = \frac{\mathbb{1}_{(instances\ C,\ s.t., f(c) > 0.5)}}{k} \%  \qquad (9)$$

where, $\mathbb{1}_{(\cdot)}$ is the indicator function which takes values 1 if the value in the parenthesis is true, else it is zero, $f$ is the trained binary classifier's output probability. The ideal value for $PVCF$ is 100%. The final $PVCF$ is the mean of the $PVCF$ values computed for input instances from the test dataset which are predicted to be in class 0.

- *PCF* is the mean feature-wise distance between counterfactuals generated with an input data instance, i.e.,

$$PCF_{cont} = -\frac{1}{k}\sum_{i=1}^{k} dist_{cont}(\boldsymbol{c_i}, \boldsymbol{x}),$$
and
$$PCF_{cat} = 1 - \frac{1}{k}\sum_{i=1}^{k} dist_{cat}(\boldsymbol{c_i}, \boldsymbol{x}),$$
(10)

where, $PCF_{cont}$ and $PCF_{cat}$ are proximity computed for continuous and categorical features, respectively and are defined as

$$dist_{cont}(\boldsymbol{c_i}, \boldsymbol{x}) \coloneqq \frac{1}{d_{cont}} \sum_{p=1}^{d_{cont}} \frac{|c_i^p - x^p|}{MAP_p},$$
and
$$dist_{cat}(\boldsymbol{c_i}, \boldsymbol{x}) \coloneqq \frac{1}{d_{cat}} \sum_{p=1}^{d_{cat}} \mathbb{1}_{(c_i^p \neq x^p)},$$
(11)

where, $d_{cont}$ is the number of continuous feature variables, $MAP_p$ is the mean absolute deviation of the $p^{th}$ continuous variable, and $d_{cat}$ is the number of categorical variables. *PCF* as defined in equation (10) can take values in the range $(-\infty, 0]$. To put a lower bound, we scale the values of *PCF* computed from different input data instances from the test dataset with the *min-max* transformation to redistribute the values between [0,1]. The final *PCF* KPI is the mean of the *PCF* values computed for input instances from the test dataset which are predicted to be in class 0. The greater the value of final *PCF* the better the KPI.

- *SCF* computes the number of changes between the original input datapoint with set of counterfactuals generated around this datapoint, i.e.,

$$SCF = 1 - \frac{1}{kd}\sum_{i=1}^{k}\sum_{p=1}^{d} \mathbb{1}_{(c_i^p \neq x^p)},$$
(12)

where, $d$ is the number of feature variables.
Like final *PCF*, the final *SCF* KPI is computed over the mean values of *SCF* computed for input instances from the test dataset which are predicted to be in class 0. The range of *SCF* is [0,1]; the closer its value to 1 the better the KPI.

- *DCF* is the pairwise distance between the set of the counterfactuals generated around an input data instance, it is defined as:

$$DCF = \frac{1}{|C|^2}\sum_{i=1}^{k-1}\sum_{j=1}^{k} dist(c_i, c_j),$$
(13)

where, $|C|$ is the cardinality of the set and $dist$ can be either $dist_{cont}$ or $dist_{cat}$. Like *PCF*, we bound *DCF* between [0,1] using the *min-max* transformation. Like the previous KPIs, the final *DCF* is the mean value of the *DCF*s from different inputs from the test dataset. The closer the value of this KPI to 1, the better it is.

Note that there exists a trade-off between *PCF* and *DCF,* and no method for counterfactual generation can maximize both these KPIs (Mothilal et al. 2020).

Table 2: RAG scores (* indicates proposed values)

| KPIs | Traffic light RAG scores | | |
|---|---|---|---|
| | Red | Amber | Green |
| 1.1.1 Presence in technical documentation | Qualitative | Qualitative | Qualitative |
| 1.1.2a VIF | > 5.0 | [1.0-5.0] | < 1.0 |
| 1.1.2b SWT | p < 0.05 | 0.05<p<0.1 | p > 0.1 |
| 1.1.2c BPT | p < 0.05 | 0.05<p<0.1 | p > 0.1 |
| 1.2.1 AUC-ROC | [0.0-0.5) | [0.5-0.8) | [0.8-1.0] |
| 1.2.2 F1-score | [0.0-0.5) | [0.5-0.8) | [0.8-1.0] |
| 1.3.1 TSVR* | [0.3-1) | [0.1-0.3) | [0-0.1) |
| 1.3.2 CSVP* | > 10 | [6-10] | [0-5] |
| 2.1.1 EqualOdds* | > 0.2 | [0.1-0.2] | < 0.1 |
| 2.1.2a DI | ≪ 0.8 | ⪅ 0.8 | (0.8-1.0] |
| 2.1.2b SP | ≫ 0.2 | ⪆ 0.2 | < 0.2 |
| 2.2.1 Diff$_{Ind}$ | ≫ 0.2 | ⪆ 0.2 | < 0.2 |
| 2.2.2 Diff$_{Ind\_GRP}$ | ≫ 0.2 | ⪆ 0.2 | < 0.2 |
| 3.1.1 $\rho_{order}$ | [-1.0-0.3) | [0.3-0.8) | [0.8-1.0] |
| 3.1.2 PUX* | > 0.2 | [0.1 – 0.2] | < 0.1 |
| 3.1.3 POIFS* | [100-20)% | [20-10)% | [10-0]% |
| 3.2.1 PVCF* | [0-75)% | [75-90)% | [90-100]% |
| 3.2.2 PCF* | [0-0.7) | [0.7-0.9) | [0.9-1.0] |
| 3.2.3 SCF* | [0-0.7) | [0.7-0.9) | [0.9-1.0] |
| 3.2.4 DCF* | [0-0.7) | [0.7-0.9) | [0.9-1.0] |

## 4. DEMONSTRATION OF THE FRAMEWORK

The audit framework presented in Table 1 is demonstrated using an open-source US health insurance dataset (Kaggle, 2018). The dataset consists of 1338 rows with age, gender, BMI, number of children, smoker, and region as independent variables and insurance charges as the dependent variable. Using binary classification, the aim is to predict if an insured is eligible to an insurance claim greater than $6,000 per region based on age, BMI, and number of children. The dataset is divided into train-test split of 95%-5%. The two classification models (MLogRM and RFM) are trained on the training dataset while the model performance metrics are evaluated on the test dataset. No hyperparameter tuning for RFM is performed. The results of the KPIs are presented in this section, which must be compared to the RAG scores depicted in Table 2.

### 4.1 Results for Statistical Properties

**RFM:** For RFM, no model assumptions need to be satisfied due to their non-parametric nature. Therefore, KPIs are not applicable for RFMs.
**MLogRM:** For MLogRM the KPI results are:
- *VIF*: VIF for age, BMI, and number of children are estimated to be 7.5, 7.8 and 1.8 respectively, implying that multicollinearity of age and BMI needs to be mitigated.
- *SWT*: The test of residuals has a *p-value* $\approx$ 0.0, implying some non-linear relationship between the independent and the dependent variables.
- *BPT*: The test for residuals has a *p-value* $\approx$ 0.0, implying lack of constant variance in residuals.

### 4.2 Results for Accuracy of Predictions

**RFM:** The of *F1*-score and *AUC-ROC* are computed to be 0.91 and 0.95, respectively on the test dataset.
**MLogRM:** The values of *F1*-score and *AUC-ROC* are 0.83 and 0.91, respectively. Both KPI values exceed 0.8 implying that this classifier possesses a good predictive and discriminatory power.

The RFM values are significantly higher than the MLogRM ones. This implies that the RFM not only has good accuracy of predictions but is also more accurate than the MLogRM. This is not surprising as RFMs are known to have higher accuracy when compared to LR models.

### 4.3 Results for Robustness

**RFM:**
- TSVR: Given RFMs do not have any model parameters, this KPI cannot be computed.
- CSVP of 0, 0, 1, 0 is computed respectively for northeast, northwest, southeast, and southwest. The value of this KPI is well within the acceptable range, implying the model output is stable around the inflection point.

**MLogRM:**
- TSVR of 0.0045, 0.0032, 0.0094, 0.0047 are computed for northeast, northwest, southeast, and southwest, respectively. The KPI values for all the regions are low, implying, the model output is not sensitive to change in model parameters (within one standard error of the estimation).
- CSVP = 1 for each of the four groups. This implies that for each of the MLM-groups there is atleast one very similar input data point pair which is predicted to be in opposite classes.

Comparing RFM with MLogRM values, we conclude that RFM is more robust for this case study.

### 4.4 Results for Discrimination

**RFM:**
- $SP = 0.01$, implying a very small difference between the sensitive groups.
- $DI = 0.98$, indicating a value well within the acceptable range of 0.8 or higher.
- $EqualOdds = 0.14$, is in the acceptable range. Note that for RFM, a region is treated as an independent categorical variable. Thus, the result presented is the average of the four regions.
- $Diff_{Ind} = 0.15$ for two very similar individuals in the region northwest, with age as the sensitive attribute and feature values (age = 30, BMI = 30, children = 2, gender = female) and (age = 35, BMI = 30, children = 2, gender = female). This value falls within the acceptable range.
- $Diff_{Ind\_GRP} = 0.26$ is computed for two individuals with same characteristics (age = 30, BMI = 30 children = 1, gender = female) in two regions, northeast and southeast. This value falls just outside the acceptable range.

**MLogRM:** Here gender (male/female) is considered to be the sensitive feature:
- $SP = 0.083$, implying a relatively small difference between the sensitive groups.
- $DI = 0.89$, indicating a value within the acceptable range of 0.8 or higher.
- $EqualOdds$ per MLM-group of 0.28, 0.5, 0.33, 0.29 for northeast, northwest, southeast, and southwest, respectively. These values are too high and therefore unacceptable.
- $Diff_{Ind} = 0.18$ for the same two individuals as in the RFM model. This is not a high difference in probability for a small difference in age.
- $Diff_{Ind\_GRP} = 0.07$ for the same two individuals as in the RFM model. The small value indicates a minor difference in fairness.

## 4.5 Results for Accuracy of Feature Attribution Explainability Methods

**RFM:** The model intrinsic feature importance method of RFM, VIM, is a global explainability method. Therefore, the feature importance values from VIM are compared with the mean of *tree*-SHAP contributions computed on the whole training dataset. Following are the KPI results:

- Mean $\rho_{order}$ of 0.93 across all regions, implying that SHAP feature contribution magnitude are highly correlated with the global model intrinsic method. However, it is important to note that model-intrinsic *tree*-SHAP has been used, which is expected to perform better than the model agnostic *kernel*-SHAP used for MLogRM.
- *PUX* for RFM cannot be computed due to the nature of VIM, where the feature contribution magnitude sums up to one.
- Mean *POIFS* = 0 across all regions. This implies that SHAP and model intrinsic feature contribution signs match perfectly.
- Feature contribution computed from the two methods, *tree*- and *kernel*-SHAP are compared for two instances from group northwest [age = 35, BMI = 40, children = 3] and [age = 35, BMI = 40, children = 2] in Figure 1. It is observed that these two methods are not aligned in magnitude, order, and sign.

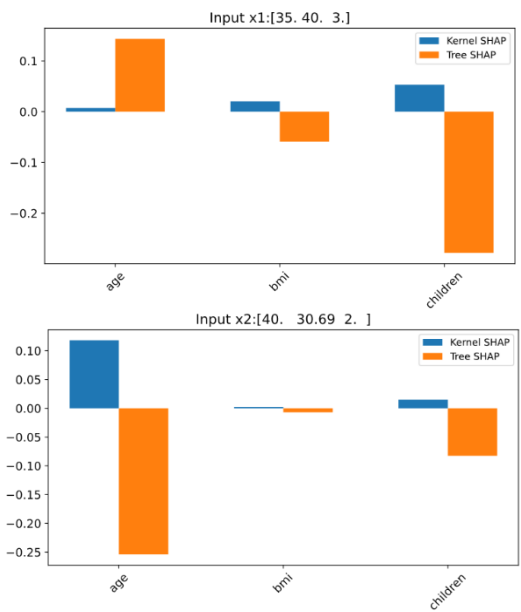

Figure 1: Comparison of feature contribution (y-axis) using tree-SHAP and kernel-SHAP methods for two data instances in the region northwest.

**MLogRM:** The model intrinsic feature importance method of MLogRM is instance-based (local). Therefore, explanations generated by this method can be compared to *kernel*-SHAP explanations per data instance. To this end, the means of $\rho_{order}$, *PUX*, and *POIFS* over 50 randomly sampled instances from the training dataset is computed and the process is repeated 10 times to estimate the KPIs for each MLM-group. The KPIs for the four MLM-groups are:

- $\rho_{order}$: $(0.86 \pm 0.07)$, $(0.80 \pm 0.05)$, $(0.80 \pm 0.09)$, and $(0.82 \pm 0.06)$. SHAP feature contribution ranks are not fully correlated with model intrinsic ones across the MLM-groups, implying that SHAP occasionally produces incorrect feature contribution magnitudes.
- *PUX*: $(0.1 \pm 0.01)$, $(0.09 \pm 0.005)$, $(0.07 \pm 0.003)$, and $(0.09 \pm 0.01)$. There exists probability gap between model intrinsic and SHAP of ~0.1 of an insured being in class 1 for all MLM-groups. This is moderately significant in this use-case.
- *POIFS*: $(9.2 \pm 1.7)\%$, $(3.1 \pm 1.04)\%$, $(7.2 \pm 2.06)\%$, and $(2.6 \pm 1.36)\%$. SHAP and model intrinsic feature contribution signs do not completely match, implying that there are instances where SHAP estimates the sign of feature contributions incorrectly.

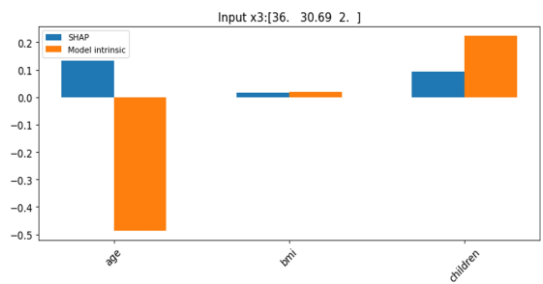

Figure 2: Comparison of feature contribution (y-axis) using kernel-SHAP and Model intrinsic methods for one data instance in the region northwest.

As demonstrated in Figure 2, kernel-SHAP fails to accurately detect the sign, magnitude and order of feature contributions for the given data instance. It is important to detect such exceptions during audit. Such examples elucidate that approximate feature attribution methods such as SHAP can produce unreliable feature explanations which are not suitable for many customer facing applications of binary classifiers.

## 4.6 Results for Counterfactual Explanations

For each data point in the test dataset for which the predicted class is 0, counterfactuals were generated. The counterfactuals were taken to be the mesh-grid of values around the data points in the test dataset such that the range of values for the feature variables are $[x_{age} - (25+rand\{0,2\}), x_{age} + (25+rand\{0,2\})]$, $[x_{bmi} - (25+rand\{0,2\}), x_{bmi} + (25+rand\{0,2\})]$, $[x_{child} - (5+rand\{0,2\}), x_{child} + (5+rand\{0,2\})]$, where $rand\{0,2\}$ is a random integer between (0,2). Note that generating counterfactuals this way is not optimal, but it is used for demonstration purposes.

**RFM:** The values computed for *PVCF, PCF, SCF,* and *DCF* are 50.28%, 0.48, 0.04, and 0.52 respectively.

**MLogRM:** The values computed for *PVCF, PCF, SCF,* and *DCF* are 33.26%, 0.53, 0.04, and 0.47 respectively.

For both RFM and MLogRM, the four KPIs, *PVCF, PCF, SCF,* and *DCF* have a Red RAG score, implying, the quality of the counterfactuals generated by the method described above is not adequate and requires improvements.

## 5. CONCLUSIONS

In this paper an audit framework for technical assessment of binary classifiers is proposed along with KPIs and corresponding RAG scores. The framework is based on three aspects: model, discrimination, transparency & explainability. The framework is demonstrated through its computed KPIs using an open-source dataset and building two commonly used binary classifiers, random forests and multilevel logistic regression. The framework suits generalized linear models more than tree-based ones. In the absence of a model intrinsic method to generate feature importance, no one feature attribution explainability method, such as SHAP is sufficient. Also, multiple KPIs are required to assess each aspect, e.g., if the KPI for *Proximity* of counterfactuals declines, the KPI for *Diversity* increases. Another example is the discrepancy in the KPIs for group fairness, *Disparate Impact* and *Equal Odds*.

Future work includes extending the framework to other classification models and launching pilots in industry. The latter is expected to provide insights, which are essential to extend the current scope of the audit framework beyond technical aspects to include organizational and process related aspects.